\documentclass[sigconf]{acmart}
\renewcommand\footnotetextcopyrightpermission[1]{} 
\acmConference[ICSE 2024]{46th International Conference on Software Engineering}{April 2024}{Lisbon, Portugal}

\usepackage{url}

\usepackage{subcaption}

\usepackage{booktabs} 
\usepackage{color} 

\usepackage{xspace} 
\newcommand{\fs}{\textsc{FuzzSlice}\xspace}

\usepackage{listings}
\usepackage{hyperref}
\usepackage{enumitem}
\usepackage{fancyhdr} 
\usepackage{tikz}
\usepackage{pgfplots}
\pgfplotsset{compat=1.18}
\usepgfplotslibrary{statistics}

\usepackage{caption}
\usepackage{subcaption}
\usepackage{algpseudocode}
\usepackage{algorithm}
\usepackage{multirow}
\usepackage{booktabs}
\usepackage{tabularx}
\usepackage{comment}
\usepackage[flushleft]{threeparttable}

\usepackage[toc,page]{appendix}

\usepackage{pdflscape}
\usepackage{afterpage}
\usepackage{rotating,graphicx}

\usepackage{makecell}

\usepackage{listings}
\usepackage{color}
\usepackage{soul}

\usepackage[skins]{tcolorbox}
\newcommand{\conclusion}[1]{\begin{center}\begin{tcolorbox}[skin=widget, left=2mm,right=2mm,top=2mm,bottom=2mm,boxrule=0.3mm,arc=0mm,coltitle=black,colframe=black!99!white,colback=white!88!gray,width=(\linewidth),before=\hfill,after=\hfill]#1\end{tcolorbox}\end{center}}
\definecolor{ScarletRed}{rgb}{0.80,0.00,0.00}

\copyrightyear{2024}
\acmYear{2024}
\setcopyright{acmlicensed}\acmConference[ICSE '24]{2024 IEEE/ACM 46th International Conference on Software Engineering}{April 14--20, 2024}{Lisbon, Portugal}
\acmBooktitle{2024 IEEE/ACM 46th International Conference on Software Engineering (ICSE '24), April 14--20, 2024, Lisbon, Portugal}
\acmPrice{15.00}
\acmDOI{10.1145/3597503.3623321}
\acmISBN{979-8-4007-0217-4/24/04}

\begin{document}



\author{Aniruddhan Murali}
\affiliation{%
  \institution{University of Waterloo}
  \state{Waterloo}
   \country{Canada}
}
 \email{a25mural@uwaterloo.ca}

\author{Noble Saji Mathews}
\affiliation{%
  \institution{University of Waterloo}
  \state{Waterloo}
   \country{Canada}
}
\email{noblesaji.mathews@uwaterloo.ca}

\author{Mahmoud Alfadel}
\affiliation{%
  \institution{University of Waterloo}
  \state{Waterloo}
   \country{Canada}
}
\email{malfadel@uwaterloo.ca}

\author{Meiyappan Nagappan}
\affiliation{%
  \institution{University of Waterloo}
  \state{Waterloo}
   \country{Canada}
}
\email{mei.nagappan@uwaterloo.ca}

\author{Meng Xu}
\affiliation{%
  \institution{University of Waterloo}
  \state{Waterloo}
   \country{Canada}
}
\email{meng.xu.cs@uwaterloo.ca}





\definecolor{codegreen}{rgb}{0,0.6,0}
\definecolor{codegray}{rgb}{0.5,0.5,0.5}
\definecolor{codepurple}{rgb}{0.58,0,0.82}
\definecolor{backcolour}{rgb}{0.95,0.95,0.92}

\lstdefinestyle{mystyle}{
    backgroundcolor=\color{white},   
    commentstyle=\color{gray},
      keywordstyle=\color{magenta},
    numberstyle=\tiny\color{codegray},
    stringstyle=\color{codepurple},
    basicstyle=\ttfamily\footnotesize,
    breakatwhitespace=false,         
    breaklines=true,                 
    captionpos=b,                    
    keepspaces=true,                 
    numbers=left,                    
    numbersep=5pt,                  
    showspaces=false,                
    showstringspaces=false,
    showtabs=false,                  
    tabsize=2,
    frame=lines,
    float=t
}

\lstset{style=mystyle}



\title{FuzzSlice: \textcolor{black}{Pruning} False Positives in Static Analysis Warnings Through Function-Level Fuzzing}

%
%
%

\begin{abstract}
Manual confirmation of static analysis reports is a daunting task. 
This is due to both the large number of warnings and the high density of false positives among them. 
Fuzzing techniques have been proposed to verify static analysis warnings. 
\textcolor{black}{However, a major limitation is that fuzzing the whole project to reach all static analysis warnings is not feasible.
This can take several days and exponential machine time to increase code coverage linearly.}

\textcolor{black}{Therefore, we propose \fs, a novel framework that automatically prunes possible false positives among static analysis warnings.
Unlike prior work that mostly focuses on confirming true positives among static analysis warnings, which inevitably requires end-to-end fuzzing, \fs focuses on ruling out potential false positives, which are the majority in static analysis reports. The key insight that we base our work on is that a warning that does not yield a crash when fuzzed at the function level in a given time budget is a possible false positive.
To achieve this, \fs first aims to generate compilable code slices at the function level.
Then, \fs fuzzes these code slices instead of the entire binary to prune possible false positives. 
\fs is also unlikely to misclassify a true bug as a false positive because the crashing input can be reproduced by a fuzzer at the function level as well.}
We evaluate \fs on the Juliet synthetic dataset and real-world complex C projects: openssl, tmux and openssh-portable. 
Our evaluation shows that the ground truth in the Juliet dataset had 864 false positives which were all detected by \fs. For the open-source repositories, we were able to get the developers from two of these open-source repositories to independently label these warnings. \fs automatically identifies 33 out of 53 false positives confirmed by developers in these two repositories. This implies that \fs can reduce the number of false positives by 62.26\% in the open-source repositories and by 100\% in the Juliet dataset.
\end{abstract}



\maketitle
\section{Introduction}
\label{sec:introduction}
Static analysis tools report errors in the source code of a program without executing it. 
These tools enable the discovery of vulnerabilities in the early stages of software development. However, they suffer from major issues. First, an overwhelming number of bugs are suggested by these tools, making it hard for a software developer to verify them, which can also lead a software development team to ignore the static analysis report~\cite{Johnson,alfadel2023discoverability}. 
Second, a static analysis tool may lack the knowledge of how data flows through the system, the dependencies and software architecture~\cite{Cheirdari,Nadeem2012Mar}.
Therefore, many of the bugs turn out to be false positives~\cite{Kang2022May,Nadeem2012Mar,Park2016May,Aloraini2017Sep}.
Such false positives produced by static analysis tools are a significant barrier to the wide-scale adoption of these tools~\cite{Christakis2016Aug,Johnson2013May}.

Fuzzing is a popular software testing technique that involves supplying arbitrary or randomized input to a computer program with the objective of uncovering unexpected behaviors, including crashes.
Prior work in fuzz testing has largely focused on identifying true positives in static analysis reports.
For example, \textcolor{black}{B{\"o}hme} et al. utilized directed grey box fuzzers to direct fuzzing towards a target location~\cite{Bohme2017Oct}. 
Other techniques use dynamic symbolic execution to verify static analysis reports~\cite{Christakis2016May}. 
However, the drawback of such approaches is the time budget and computational power required to reach all static analysis warnings~\cite{Bohme2020Nov}.
{In this paper, we propose \fs, an approach that aims to prune false positives produced by static analysis tools.
The novelty of \fs is twofold: 
\begin{itemize}
    \item \textbf{Conceptual Innovation:} While several other techniques (e.g., ~\cite{Bohme2017Oct, Taintscope, 272157, defuzz}) aim at identifying true positives in static analysis warnings, \fs is optimized towards pruning possible false positives in a given program. 
Instead of fuzzing the entire program from the main function, \fs only fuzzes the function containing the warning to prune possible false positives.
\fs hinges on the novel idea that a flagged code fragment (represented as a warning) executed at least once at the function level and not yielding a crash in a given time budget is a possible false positive.\\

    \item \textbf{Technical Innovation (close-to-warning fuzzing):} Unlike typical methods that reduce fuzzing cost by independently fuzzing modules and libraries~\cite{Wei2022May,Jang2019Dec,Somorovsky2016Oct,Chen2022Nov}, \fs focuses on generating compiled slices that encompass the warning location detected by a static analysis tool.
\end{itemize}

\fs generates and fuzzes a separate binary for each warning, which facilitates the coverage of most warnings with reduced computational cost (under 5 minutes of fuzzing). Finally, \fs is unlikely to misclassify a true bug as a false positive because the crashing input can be reproduced by a fuzzer at the function level.
}

We evaluate \fs on four diverse repositories comprising one synthetic and three open-source datasets that have been reported to contain buffer overflow vulnerabilities.
Our evaluation shows that the ground truth in the synthetic Juliet dataset has 864 false positives which were all confirmed by \fs. 
For open-source repositories, we found 143 possible false positives among 265 warnings. We reached out to developers from two of these open-source repositories (tmux and openssh-portable) to label these warnings. \fs automatically identifies 33 possible false positives out of 53 false positives confirmed by developers in these two repositories.
%
%
%

In summary, this paper makes the following key contributions:
\begin{itemize}
    \item We introduce  \fs, a novel design built upon the insight that warnings fuzzed at the function level and not resulting in crashes within a reasonable time budget are possible false positives.
    \item  \fs efficiently identifies possible false positives in static analysis reports by: (1) automatically generating a minimal compiled code slice for complex real world C code encapsulating \emph{any} arbitrary static analysis warning, and (2) generating a fuzzing wrapper that performs type-based input generation for the function enclosing the warning.
    \item We develop a prototype tool for \fs. The tool and datasets along with a docker image are publicly available \footnote{\url{https://archive.softwareheritage.org/browse/origin/https://github.com/NobleMathews/FuzzSliceICSE}}
\end{itemize}

\section{Motivational Example}
 \label{sec:motivation}
The goal of this study is to \textcolor{black}{prune possible false positives} efficiently. 
In this section, we provide an example to motivate the \fs approach. An example code of the openssl repository~\cite{openssl2023Jan} in the C language is shown in Listing~\ref{Motivating example}. The code listing describes a function (i.e., \verb|glue_strings|) that joins an array of strings (the function argument) into a single string (the return value). 
On line 10, the variable \verb|len| is updated in a loop to hold the sum of the length of all input array strings. 
On line 12, the variable \verb|ret| is dynamically allocated with a size of \verb|len+1|. 
On line 16, the string is joined together in pointer variable \verb|ret| by iterating the pointer \verb|p| and copying each input string one by one. 

When a static analysis tool such as RATS~\cite{andrew-d.2023Jan} is run on this code it flags line 16 as a possible heap buffer overflow. However, this is clearly a false positive because the \verb|strcpy| on line 16 can never exceed the bounds of the allocated pointer \verb|ret|. The reason behind this is that the size of the allocated pointer \verb|ret| will always be one plus the length of all input strings. RATS is not capable of this kind of value flow analysis for variable \verb|len|. Therefore the tool cannot be sure that line 16 will never cause a heap buffer overflow.

Fuzzing has become a popular solution for the verification of static analysis reports. It is possible to compile the whole openssl binary and guide the fuzzing toward line 16 in the code snippet from the \verb|main| method. However, this often takes several days, \textcolor{black}{many CPU cycles} and requires the help of appropriate fuzzing dictionaries. In spite of this, there is no guarantee that the given static analysis warning can be covered by fuzzing within a given time budget.

\fs is an approach that is primarily targeted toward false positives in static analysis warnings. It takes advantage of the fact that the function \verb|glue_strings| in Listing~\ref{Motivating example} can be fuzzed directly to identify it is a false positive. Given an arbitrary warning location, it automatically constructs a function slice of the program, compiles it including necessary dependencies, generates its own fuzzing wrapper and fuzzes the function slice. When the function slice is fuzzed on its own, the static warning on line 16 can be easily reached. Let us assume that fuzzing this function slice gives no crash on line 16. This implies that fuzzing from \verb|main| will also not result in a crash. This can be derived from the fact that caller functions can only constrain the input to a given function through the function arguments. On the other hand, if a crash is observed on the static analysis warning line then we cannot comment on it being an actual bug or a true positive. This is because a caller function can invalidate the input that causes the observed crash in the slice. 
\fs aims to \textcolor{black}{prune all false positives} within a static analysis report similar to Listing~\ref{Motivating example}.

\begin{lstlisting}[language=C++, caption=Code snippet from the openssl project flagged by RATS as buffer overflow~\cite{exampleloc}., label={Motivating example}]
 /* Glue an array of strings together and return it as an allocated string.
 */
char *glue_strings(const char *list[])
{
    size_t len = 0;
    char *p, *ret;
    int i;

    for (i = 0; list[i] != NULL; i++)
        len += strlen(list[i]);

    if (!(ret = p = OPENSSL_malloc(len + 1)))
        return NULL;
        
    for (i = 0; list[i] != NULL; i++)
        p += strlen(strcpy(p, list[i])); //False positive

    return ret;
}
\end{lstlisting}

\section{FuzzSlice Approach}
\label{sec:approach}

We split our approach into two parts. First, we discuss the main design steps of \fs.
Then, we discuss in detail how we achieve each step.


\subsection{Design}
The core functionality of \fs is to decide \emph{whether a specific static analysis warning is a possible false positive} (and hence, can be de-prioritized for manual triage).
More formally, given a warning $w$ in a static analysis report for program $P$, \fs examines $w$ in three conceptual steps, namely (D1) Minimal Slice Creation; (D2) Fuzzing Input Generation; and (D3) Warning Classification.\\

\noindent
\textbf{(D1) Minimal Slice Creation.}
First, we build an execution environment that fully encloses $w$ at the function level. This is called a slice of the original program denoted by $S$. Ideally, $S$ should have the following properties:
    \begin{enumerate}[label=(\alph*)]
        \item $S \subseteq P$, i.e.,
        the slice should be smaller than the program unless
        the program is just a single main function.
        \item Consider the function $F$ directly enclosing warning $w$. This function must be part of the slice i.e., $F \subset S$
        \item Consider a function $F_2$ which is called by a function $F_1$ within slice $S$ ($F_1 \subset S$). In that case, $F_2$ is also part of the slice $S$ i.e., $F_2 \subset S$.
    \end{enumerate}

\noindent

 By definition of property (c), if a function $F_k$ is not called by \textit{any} function $F_i$ within $S$ then $F_k$ cannot be in $S$. This implies that any execution beginning from $F$ (the function enclosing the warning) can never reach $F_k$. Therefore the defined slice $S$ is a \emph{minimal} slice capturing execution environment related to warning $w$ at the function level.\\
 

\noindent
\textbf{(D2) Fuzzing Input Generation.} Next, \fs generates valid and versatile inputs with the goal of testing the minimal slice $S$ comprehensively. \fs achieves this with randomly generated arguments used by the function $F$ which encloses the warning. \fs performs type-based input generation and mutation, i.e., unlike AFL~\cite{BibEntry2023Mar} or libfuzzer~\cite{BibEntry2023Jan} which blindly mutates raw bytes, the input generator in \fs recognizes the type of the arguments, including struct pointers and mutate inputs based on typing rules. 
\textcolor{black}{This helps prevent early-termination by input sanitization logic during the execution of F which improves both the efficiency and effectiveness of fuzzing the slice.}\\


\noindent
\textbf{(D3) Warning Classification.} Finally, we decide whether $w$ is a possible false positive, i.e., based on fuzzing the minimal slice $S$:

    \begin{enumerate}[label=(\alph*)]
        \item If the fuzzer finds a concrete input $I$
            that causes a dynamic bug checker (e.g., ASAN)
            to report an issue on $w$ while executing $S$,
            it is possible that either this confirms $w$ to be
            a true positive (hence, higher priority for manual triage)
            or that $I$ is an infeasible input which can never be generated
            when executing from the `main` function.
        \item If the fuzzer cannot find an input that
            causes a dynamic bug checker to complain about $w$
            even after exhausting its computation budget and executing $w$ at least once, it is possible that $w$ is a false positive based on the fact that even with a free-form search, the fuzzer cannot find an offending input to
            trigger the warning.
    \end{enumerate}

\noindent
Implementing the design steps above presents unique challenges. We discuss how we tackle these challenges in the next section.

\subsection{Proposed technique}
\label{technique}
\graphicspath{ {./images/} }
\begin{figure*}[tb!]
    \centering
\includegraphics[width=0.85\linewidth]{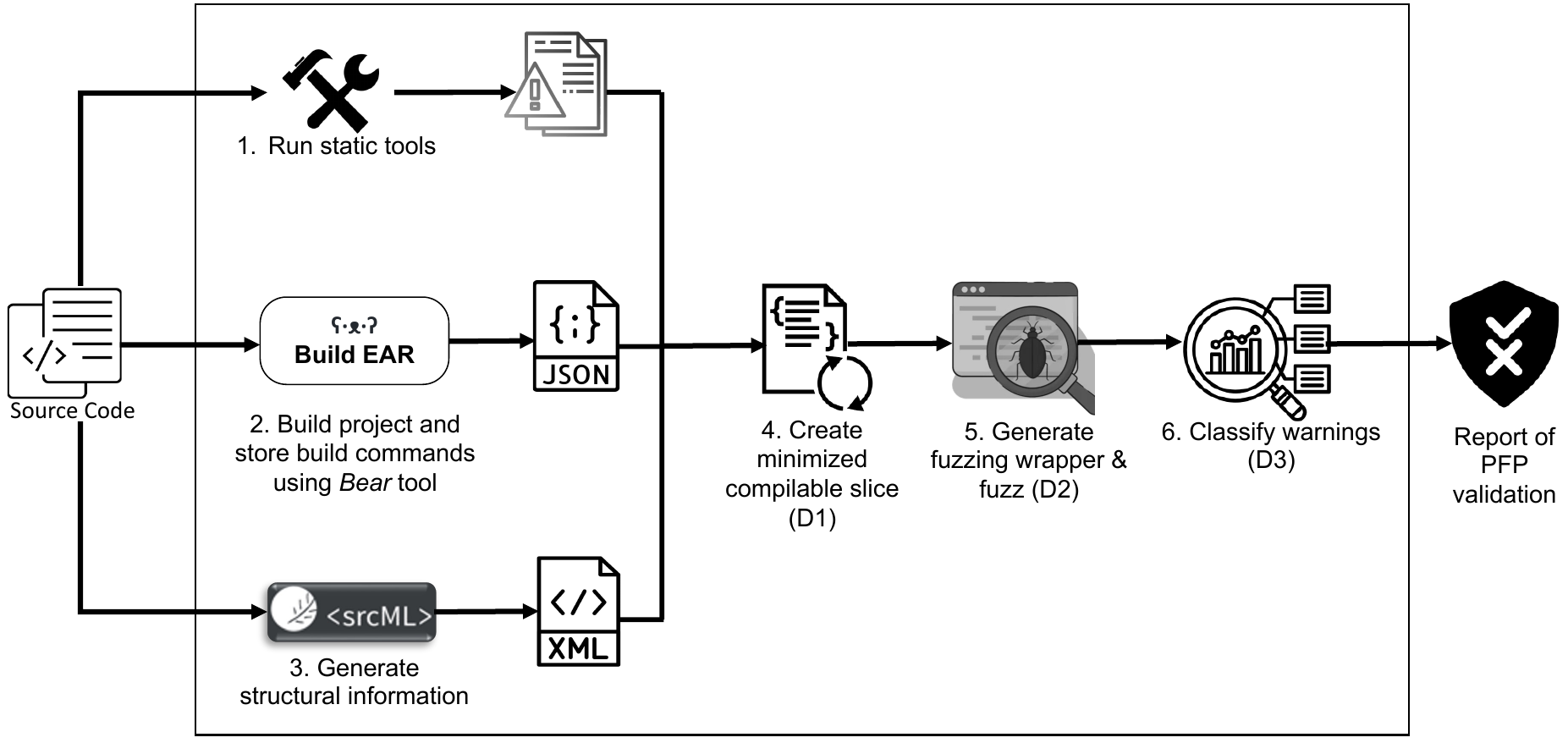}    \caption{Overview of the \fs technique.}
    \label{Approach}
\end{figure*}

In this section, we discuss the technical aspects of each of the design components in the previous section. An overview of \fs implementation is shown in Figure~\ref{Approach}.

\fs is used when a software developer
would like to validate and prioritize a set of warnings from a static analysis tool.
\fs takes the following steps to rule out potential false alarms in these static analysis warnings: (1) Build project and store build commands;
(2) Generate structural information of source code in XML format;
(3) Create a minimized compilable slice containing the warning;
(4) Generate fuzzing wrapper; and
(5) Classify warnings into possible false alarms
(remaining are worth further investigation).

%
%
Along this process,
step (1) and (2)
are information collection steps that aim to
build a compiler-agnostic representation of the
program source code;
while steps (3), (4), and (5)
outline a concrete implementation to achieve
each of the three design goals (D1), (D2), and (D3)
respectively.


We first describe each step in detail using the code example in Listing~\ref{Motivating example}.
Next, we walk through each step to show how the given code example is finally classified by \fs to be a possible false positive.
%

In the example of Listing~\ref{Motivating example}, the static analysis tool flags the \verb|strcpy| on line 16 as a possible heap buffer overflow. \fs currently uses two static analysis tools RATS~\cite{andrew-d.2023Jan} and Infer~\cite{Calcagno2011} to generate static analysis reports. 
It is important to note that \fs can use any static analysis tool in principle for this approach.\\

\noindent
\textbf{Step 1: Build the project and store build commands.} 
In the first step of our approach,
\fs performs a  build of the repository using its native build system (usually \textsc{make} or \textsc{cmake} for C-based projects). During the build process, \fs stores important build information (e.g., include paths for header files, compiler options, shared library locations, path to compiled file etc.), which will be used later to compile minimized slice $S$.
\fs utilizes Build EAR - a tool that generates a compilation database for a given build process~\cite{rizsotto2023Jan}.
Build EAR stores build commands related to each file in a JSON format.  In our running example, the code snippet lies in a file called \verb|driver.c|. The build command of \verb|driver.c| is stored in JSON format in our example. 
\fs uses the generated JSON files when compiling the code slice in Step 3.

Original source code can have complex structures such as functions which are generated dynamically using macros or preprocessor directives that allow multiple definitions of a function for different operating systems. To alleviate this, we use the first stage of compilation to preprocess the source code files. This is necessary because it makes the code easier for minimization while generating code slices in later steps.
The preprocessed files are similar to original source code but stripped of all comments, have inline macros substituted, and preprocessor directives like \verb|#ifdef| removed.
We achieve this step by adding some flags to the compilation process (e.g., ``\verb|-save-temps|'' flag).
 In the example of our running case in Listing~\ref{Motivating example}, only the comment gets stripped after preprocessing.\\

\definecolor{eclipseStrings}{RGB}{42,0.0,255}
\definecolor{eclipseKeywords}{RGB}{127,0,85}
\colorlet{numb}{magenta!60!black}

\lstdefinelanguage{json}{
    basicstyle=\normalfont\ttfamily,
    commentstyle=\color{eclipseStrings}, 
    stringstyle=\color{eclipseKeywords}, 
    numbers=left,
    numberstyle=\scriptsize,
    stepnumber=1,
    numbersep=2pt,
    showstringspaces=false,
    breaklines=true,
    frame=lines,
    string=[s]{"}{"},
    comment=[l]{:\ "},
    morecomment=[l]{:"},
    literate=
        *{0}{{{\color{numb}0}}}{1}
         {1}{{{\color{numb}1}}}{1}
         {2}{{{\color{numb}2}}}{1}
         {3}{{{\color{numb}3}}}{1}
         {4}{{{\color{numb}4}}}{1}
         {5}{{{\color{numb}5}}}{1}
         {6}{{{\color{numb}6}}}{1}
         {7}{{{\color{numb}7}}}{1}
         {8}{{{\color{numb}8}}}{1}
         {9}{{{\color{numb}9}}}{1}
}


\noindent
\textbf{Step 2: Generate structural information of source code in XML format.} In this step, we aim to label AST-like high-level structural information within the entire project repository. 
This AST-like information will be used to search for code sections that represent relevant functions to add to minimized slice $S$ in the next step. 
To achieve this, we use srcML, a tool that provides an XML format for structural information of source code~\cite{Collard}. 
It is lightweight and highly scalable. Also, the output of srcML (XML-format) makes it easier when parsing and searching for C references (functions or variables) while constructing minimized compilable slice.
\fs obtains the srcML of the preprocessed files (obtained in Step 1). 
The XML produced by srcML labels individual nodes that represent source code components (e.g., functions, declaration statements, for loops etc.). 
At the end of this step, we have XML equivalent of source code with all high level structural information labelled, which we use in Step 3.\\






\algnewcommand\algorithmicforeach{\textbf{for each}}
\algdef{S}[FOR]{ForEach}[1]{\algorithmicforeach\ #1\ \algorithmicdo}

\begin{algorithm}[tb!]
\caption{Minimal Slice Creation}\label{pseudocode}
\begin{algorithmic}[1]
\State \textbf{Input:} Function F containing the warning
\State \textbf{Output:} Slice S
\Procedure{Slice}{Dependency F}

\State $\textit{file} \gets \textit{EnclosingFile}(F)$
\State $\textit{filesrcML} \gets \textit{GetSrcML}(file)$
\State
\State $\textit{queue} \gets \textit{F}$ \Comment{File level breadth-first search}
\ForEach {$c \in  queue $} 
    \If{$c \in \textit{GetFunctions}(filesrcML)$}
        \State $\textit{S} \gets \textit{S} \cup \textit{c}$
        \State $\textit{queue} \gets \textit{queue} \cup \textit{GetCallees}(c, filesrcML)$
        \State $\textit{Pop } \textit{c} \textit{ from } \textit{queue}$
    \EndIf
\EndFor

\State

\State $extDependencies \gets CompileSlice(S)$ \Comment{Recursion}
\ForEach {$c \in extDependencies$}
    \State \Call{Slice}{c}
\EndFor
\EndProcedure
\end{algorithmic}
\end{algorithm}

\noindent
\textbf{Step 3: Constructing minimized compilable code slice.}
In this step, \fs obtains a compilable code slice that contains a given warning, i.e., the code slice that \fs generates comprises the entire function that encapsulates the warning and its dependencies in the same file as well as in other files. In the design section, we defined dependencies as function callees. In this section, we generalize the concept of dependency to a reference that includes functions, structs, or global variables required for successful code slice compilation. In our running example, \verb|OPENSSL_malloc| is a dependency of the function \verb|glue_strings|. The function  \verb|glue_string| exists in a file called \verb|driver.c| while its dependency \verb|OPENSSL_malloc| exists in another file called \verb|alloc.c| in the openssl project.

Within the design section, we discussed properties (a), (b) and (c) of minimized slices in (D1). Here we enforce these different properties of the minimized slice. To enforce these properties, we utilize the stored build commands (from Step 1) and source code srcML outputs (from Step 2).

First, we enforce property (b) of the minimized slice in the design section (D1) by identifying the function $F$ enclosing the warning and adding it to slice $S$. For our example in Figure ~\ref{Motivating example}, the enclosing function is \verb|glue_strings| which can be identified by parsing the srcML. 
This function is the first function added to the code slice.


Now, we try to enforce properties (a) and (c) in (D1) while creating minimized slice $S$. We recursively identify all other dependencies required by the function enclosing the warning in all files across the repository. We collect all the required dependencies (including callee functions) automatically over several iterations using compiler logs.
To achieve that, we perform the following steps: (1) code minimization within the file, (2) Attempt to compile; if unsuccessful - identify other files containing the missing dependencies (3) Recurse until compilation in step (2) succeeds. We describe the pseudocode of our steps in algorithm~\ref{pseudocode}. We explain the pseudocode as we describe each step in detail.\\

\noindent
(1) \textit{Code minimization within a file.} In this step, we retain all dependencies needed by a given function (e.g., \verb|glue_strings|) within its file.
To do this, we obtain the file containing the required function which initially is the file containing the warning. 
We automatically recurse the callees of the given function in a breadth-first search manner. We use srcML labelled nodes (from Step 2) to identify function calls within the given file. These dependencies are retained during the minimization process. 
Finally, we filter unused dependencies (not covered by the breadth-first search) which are not relevant to the code slice that \fs is trying to create in this file. We shall henceforth call this file the minimized file. Lines 4-14 in algorithm~\ref{pseudocode} create the minimized file through a breadth-first search before adding to the slice $S$.
For the code in listing~\ref{Motivating example}, the file to be minimized is \verb|driver.c|. Only the function \verb|glue_strings| is retained within \verb|driver.c|. \\

\noindent
(2) \textit{Attempt to compile.} 
We use the compile commands (stored by Build EAR in Step 2) for the given minimized file. 
Using these compile commands, we attempt to compile and obtain the object files. In the case that the compilation is successful, the compilable code slice is ready for the next step (i.e., generating a fuzzing wrapper). 
If the build fails, we automatically parse the compiler error logs to identify missing references. These missing references can be external dependencies in other files that we have not minimized yet. This step is shown in line 16 in algorithm~\ref{pseudocode}, where the external dependencies are obtained from the compiler logs.
In our running example, \verb|OPENSSL_malloc| is a missing reference thrown as an error by the compiler, as shown in Listing~\ref{compiler-error}. Hence, \fs searches for this reference among other files in the srcML representation of the repository and locates it in another file \verb|alloc.c|.

\begin{lstlisting}[language=C++, caption=Compiler error requesting additional references., label={compiler-error}]
driver.c:12:19: error: implicit declaration of function 'OPENSSL_malloc' is invalid
  if (!(ret = p = OPENSSL_malloc(len + 1)))

\end{lstlisting}

\noindent
(3) \textit{Recurse until compilation succeeds. } Previously, we attempted to compile the minimized file. Only when we fail, we recurse over new required references. In our running example, \verb|OPENSSL_malloc| is the new required reference and \verb|alloc.c| is the new file that must be minimized. This recursion step is shown in lines 17-19 in algorithm~\ref{pseudocode}.
We repeat (1) with \verb|OPENSSL_malloc| as the required function and \verb|alloc.c| as the file to be minimized.  After this recursion, our running example will become a compilable code slice.

By the end of this step, we obtain all references within multiple minimized files required for the function enclosing the warning $F$. This is the minimized slice $S$ described in the design section. Additionally, we have also successfully compiled these files. Finally, we have a list of object files which will be linked  along with a fuzzing wrapper in the next step.\\

\noindent
\textbf{Step 4: Generate fuzzing wrapper. }  In this step, we generate a fuzzing wrapper tailored to each function that contains a given static analysis warning. This step aims to generate versatile inputs to fuzz the minimized slice $S$ according to (D2). We require a fuzzing wrapper which is a piece of code that will correctly initialize the arguments to this function and all its fields (based on type) so that we can reach the warning through fuzzing.

To create the fuzzing wrapper, we write a Python script that looks at the argument type and initializes it correctly depending on the argument type. For primitive C types (eg. char, int, bool, double etc.), the fuzzing wrapper handles each case in a unique way that appropriately fuzzes them.  Within our example, the function \verb|glue_strings| has only one function argument \verb|char** list| (list of strings) that needs to be fuzzed. The generated fuzzing wrapper for the example is shown in Listing~\ref{Fuzzing wrapper}. It has two inputs \verb|Fuzz_Data| which is the fuzzing bytes and \verb|Fuzz_Size| which is the length of these fuzzing bytes. The fuzzing wrapper uses these fuzzing bytes to randomly initialize the argument \verb|list|. For this purpose on line 7 the first few fuzz bytes are read which is used to split the fuzz bytes into chunks on line 10. The wrapper then iterates in a for loop allocating all the strings on line 16 and copying the fuzz bytes on line 18. This fuzzed function argument \verb|list| is then passed to the function \verb|glue_strings|.

There can also be other cases where user-defined structs or objects are passed as arguments to the function to be fuzzed. These objects can have their own fields which must be correctly initialized. In this case, we use GDB - a debugger for C~\cite{stallman1988debugging} to resolve the object into the C primitive types automatically. In the fuzzing wrapper, \fs initializes the fields appropriately within the object and finally fuzzes only the primitive types.
\begin{lstlisting}[language=C++, caption=Fuzzing wrapper., label={Fuzzing wrapper}]
int LLVMFuzzerTestOneInput(uint8_t* Fuzz_Data, size_t Fuzz_Size)
{
    uint8_t * pos = Fuzz_Data;
    // Use fuzz bytes to find no. of strings
    char  **list;
    size_t num_ptr;
    memcpy(num_ptr, pos, sizeof(size_t));
    num_ptr = 1 + abs(num_ptr) % Fuzz_Size;
    // Find length of each string from fuzz bytes
    size_t str_size = Fuzz_Size/num_ptr; 
    // Allocate pointers first
    list = malloc(num_ptr * sizeof(char*));
    for (int i=0; i< num_ptr; i++ )
    {
        // Allocate string
        list[i]= malloc(str_size);
        // Copy fuzzed characters
        memcpy(list[i], pos, str_size);
        pos += str_size;
    }
    // Call target function
    glue_strings(list);
    //Free allocated variables after this
}
\end{lstlisting}

\fs has now generated a fuzzing wrapper for the code slice that finally calls the function enclosing the warning. Once the fuzzing wrapper is in place, the fuzzing wrapper is compiled. Then all the object files from Step 3 and the fuzzing wrapper are linked together using the link commands from Build EAR stored in Step 1. We then use ASAN~\cite{serebryany2012addresssanitizer} as an oracle which crashes the program during stack and heap buffer overflows during the fuzzing process.
At the end of this step, we have a compiled binary ready to be fuzzed. We then use \textsc{libfuzzer}~\cite{BibEntry2023Jan} for the purpose of fuzzing.\\

\noindent
\textbf{Step 5: Classify warnings. } 
This step is aimed at tackling the classification of the warnings after fuzzing the minimal slice $S$ (D3).
At this stage of our approach, each static analysis warning has its own binary which is compiled and linked. 
\fs fuzzes each binary after which the llvm-coverage~\cite{BibEntry2023Jan} is obtained to show the number of times each line is executed during fuzzing.
We classify the output of fuzzing the binary in only one of four states as follows:

\begin{enumerate}
    \item There is at least one crash/buffer overflow at warning location - \textbf{Crash (C)}
    \item There is no crash/buffer overflow at the warning location, but the line is executed - \textbf{Possible False Positive (PFP)}
    \item The warning line is not executed - \textbf{Not Reachable (NR)}
    \item The slice is not compiled - \textbf{Not Compiled (NC)}
\end{enumerate}

When there is no crash or buffer overflow at the warning location but the line is executed according to coverage, then we can predict that the warning has a high chance of being a false positive. This is because a caller of the function enclosing the warning can only constrain the function argument values compared to the fuzzer. However, in the case of a crash, we cannot confirm that the warning is a true positive because the caller function may invalidate the crashing input. Similarly, if a given line is not executed according to coverage or if a code slice fails to compile, we cannot say anything about the warning.\\

\noindent
\textcolor{black}{\textbf{Novelty of the approach:} The novelty of the \fs approach lies in its ability to generate compiled slices for fuzzing. This means that for a static analysis warning anywhere in the repository, the \fs framework automatically identifies the required dependencies of the enclosing function, compiles, and links the slice with the correct compiler options. We generate a unique binary aimed at covering each warning. We believe this is a novel idea within the \fs framework especially when combined with fuzzing wrapper generation to prune possible false positives.
}

In the next section, we evaluate the \fs approach and discuss resulting classes, with a focus on minimizing false positives. \fs aims to assist developers in efficiently de-prioritizing false positives without extensive manual effort.


\section{Evaluation}
\label{sec:results}
In this section, we evaluate our proposed approach.
\textcolor{black}{In particular, we aim to answer the following research questions in our evaluation:}

\begin{itemize}
\item \noindent
\textcolor{black}{\textbf{RQ1.} How many PFPs can \fs confirm on a synthetic dataset?}

    \item \textcolor{black}{\textbf{RQ2.} How many PFPs can \fs confirm on a real-world project dataset?}
    \item \textcolor{black}{\textbf{RQ3.} How does  \fs perform in terms of coverage, warning executions and compilation for PFP warnings?}
\end{itemize}


In order to answer the above research questions, we first introduce our evaluation setup in Section~\ref{setup}.
Then, we present our evaluation results of each research question in Section~\ref{res1}, Section~\ref{res2}, and Section~\ref{res3}.


\subsection{Evaluation Setup}
\label{setup}
In this section, we describe the static analysis tools and the benchmarks used to evaluate \fs.\\

\noindent
\textbf{Static analysis tools.} We use two static analysis tools, RATS~\cite{andrew-d.2023Jan} and Infer~\cite{Calcagno2011}. 
RATS is an open-source tool that utilizes a vulnerability database to flag similar code as a warning. 
RATS detects buffer overflows and race conditions.
The second tool Infer is developed and used internally by \textcolor{black}{Meta}. 
Infer performs abstract interpretation that reasons about mutations to computer memory to detect buffer overflows and null dereferences. Both of these tools output warnings at different severity levels. We use the "High" and "Medium" severity of warnings for RATS and L1 and L2 severity of warnings for Infer because they are the most faithful warnings for these tools.
Both static analysis tools provide the warning at the line level within a given file. 
For our evaluation benchmark, we obtain two sets of warnings, one from each tool.\\

\noindent
\textbf{Datasets.} We use two datasets to evaluate \fs. We are interested in \textcolor{black}{pruning false positive buffer overflow warnings} in both datasets. 
First, we use a synthetic benchmark called \textit{Juliet test suite}~\cite{Boland2012Oct}.
We run \fs on Juliet test suite v1.2 for C/C++, which is a benchmark created by the US National Security Agency (NSA) specifically for assessing the capabilities of static analysis tools. The benchmark labels each possible warning location with comments to indicate that the warnings are true or false positives.

Second, we evaluate \fs on three real world open-source repositories, namely, \textit{openssl, tmux, and openssh-portable}~\cite{openssl2023Jan,openssh2023Jan,tmux2023Jan}. 
The selected packages for evaluation represent various domains. Openssl is a robust, commercial-grade toolkit for the Transport Layer Security (TLS) protocol. 
Tmux is an open-source multiplexer for Unix-like operating systems. Using tmux, multiple  terminal sessions can be accessed in a single window. 
Openssh is the primary connectivity tool for remote connectivity through ssh protocol eliminating eavesdropping and hijacking.
Table~\ref{Repo stats} presents descriptive statistics on the selected packages. The data shows that these repositories are actively maintained and have a large number of lines of code.
\textcolor{black}{The following are the git versions of each repository used for the analysis:} openssl (894f2166ef), tmux (70ff8cfe), and openssh-portable (5f93c483).

\textcolor{black}{In selecting datasets for our evaluation, we opted to consider both synthetic and open-source repositories. 
Our reasoning for considering both synthetic and open-source repositories stems from the observation that synthetic benchmarks, such as Juliet, tend to yield a higher ratio of true positives to false positives, making them an effective means of assessing the performance of \fs on a large number of true positives. 
Additionally, a synthetic dataset provides ground truth, which can be used to \emph{objectively} evaluate the effectiveness of \fs.
In contrast, static analysis warnings obtained from open-source repositories are likely skewed toward false positives. 
Hence, we believe that by considering both synthetic benchmarks and open-source repositories, we can obtain a more comprehensive and reliable evaluation of \fs performance.}\\

\noindent
\textbf{\fs Configuration.} 
To evaluate the warnings produced by the static analysis tools, we subject each warning to a fuzzing process lasting five minutes. 
Our fuzzing procedure was conducted on a Headless Server equipped with a powerful hardware configuration consisting of 64 cores of Intel Xeon Gold 6226R processors, operating at a speed of 3.900GHz, and 128GB of RAM on Ubuntu 20.04 LTS.
It is worth noting that certain warning locations may only be compilable on specific operating systems as defined by preprocessor directives. As a result, \fs may not be able to generate a compilable code slice for these warning locations due to the lack of build information. Therefore, we excluded such warnings from our analysis.

\begin{table}[tb!]
   \small
  \caption{Statistics on the three project repositories in our benchmark.}
  \centering
 \resizebox{0.8\columnwidth}{!}{ 
 \begin{tabular}{l r r}
    \toprule
    {Repository} & {Lines of code} & {Latest commit}\\
    \midrule
    openssl & 450,982 &29/03/2023\\ 
    tmux & 106,528 &15/03/2023 \\ 
    openssh-portable & 60,387 & 29/03/2023 \\ 
  \bottomrule
\end{tabular}
}
\label{Repo stats}
\end{table}


\begin{table}[tb!] 
  \caption{\fs performance on Juliet test suite.}
  \centering
 \resizebox{0.77\columnwidth}{!}{ 
 \begin{tabular}{l r r r r r}
    \toprule
    \centering
    {Ground Truth} & 
    {Total} & {PFP} &
      {C} & {NR} & 
    {NC}\\
    \hline
    True positive & 1,059 & 20  & 1,039 & 0 & 0\\
    False positive & 864 & 864 & 0 & 0 & 0 \\
    \midrule
    Total & 1,923 & 884 & 1,039 & 0 & 0\\
  \bottomrule
\end{tabular}
}
\label{julietstats}
\end{table}

\subsection{RQ1: Synthetic Dataset (Juliet Test Suite)}
\label{res1}
The goal of evaluating \fs on a synthetic data set
(which provides ground truth)
is to validate the following foundational insights behind \fs - a false alarm by a static analysis tool should not be flagged by the dynamic checker in \fs while a true bug can and is likely to be caught by the dynamic checker in
\fs as well.

Our evaluation involved the use of version 1.2 of the Juliet test suite, which provides ground truth for all static analysis warnings, thereby facilitating their classification as either true or false positives.
We run RATS and Infer over the Juliet dataset, resulting in a total of 1,923 unique static warnings which comprises of 864 false positives and 1,059 true positives. Subsequently, these warnings were subjected to fuzzing using \fs.

Table~\ref{julietstats} shows that \fs identified all of the 864 false positives within the static analysis warnings. It is worth noting that \fs was able to execute all of the warning lines classified as false positives without observing any crashes. 
This suggests that \fs has the potential to effectively \textcolor{black}{prune false positives} in a static analysis report.

In addition, \fs was able to compile and execute all of the static analysis warnings, including 1,059 true positives that suggest possible buffer overflow vulnerabilities. 
Of the total 1,059 warnings labelled as true positives, indicating a possible buffer overflow at the warning line number, \fs was able to crash the warning line in 1,039 cases.
However, there were only 20 instances in which \fs was unable to crash the warning line due to the involvement of global variables. 
In these 20 instances, \fs wrongly classified them as false positives.
Since \fs does not currently fuzz these global variables, default values with which they are initialized were used instead. We discuss more about this in Section~\ref{sec:futurework}.


\conclusion{
    \textbf{Answer to RQ1:}
   We find that \fs is able to compile minimized slices for all warnings in the Juliet dataset. Out of 864 false positive warnings in the Juliet dataset, \fs confirms all of them by executing the warning without observing any crashes.
}

\subsection{RQ2: Real-World Dataset}
\label{res2}

We evaluate \fs on real-world open source projects with the goal of assessing practicality of \fs in handling large codebases. We also would like to see if developers agree with the labelling provided by \fs.


We evaluate \fs on three popular open-source repositories: openssl, openssh-portable and tmux~\cite{openssl2023Jan,openssh2023Jan,tmux2023Jan}. 
Our results are presented in Table~\ref{openstats}. As shown in the table, \fs was able to \textcolor{black}{prune} 143 instances of possible false positives (PFP) (54\%) out of 265 warnings. This means that \fs executed 143 static analysis warnings without any observed crashes at the warning location. Additionally, \fs detected 25 crashes (9.4\%) at the warning location out of the 265 warnings. It should be noted that these crashes may be false positives if the callers of the enclosing function invalidate the crashing inputs. However, these warnings can still be useful for developers to prioritize for further manual triage.
Lastly, we point out that \fs detected 76 warnings that were not reachable, and encountered 21 warnings (8\%) for which it could not generate a compiled slice. The relatively low number of not compiled cases (i.e., 21) is a testament to \fs ability to minimize complex code with features such as macros, function pointers, and structs.
This demonstrates that \fs is capable of minimizing and compiling a wide range of real-world, complex C code.
We delve into the reasons behind the not reachable and not compiled cases in Section~\ref{sec:futurework}.\\

\begin{table}[tb!]
  \caption{\fs performance on the three studied open source repositories.}
  \centering
 \resizebox{0.85\columnwidth}{!}{ 
 \begin{tabular}{l l r r r r r}
    \toprule
    {Repository} & {Tool} & {Total} & {PFP} & {C} & {NR} & {NC}\\
    \midrule
\multirow{2}*{openssl} & RATS &  30 & 21 & 1 & 8 & 0\\
& Infer & 163 & 88 & 18 & 45 & 12\\
\hline
\multirow{2}*{tmux} & RATS & 5 & 4 & 1 & 0 & 0\\
& Infer & 18 & 6 & 2 & 10 & 0\\
\hline
\multirow{2}*{openssh-portable} & RATS & 20 & 6 & 2 & 10 & 2\\
& Infer & 29 & 18 & 1 & 3 & 7\\

\hline
Total &  & 265 & 143 & 25 & 76 & 21\\
\bottomrule
\end{tabular}
}
\label{openstats}
\end{table}







\noindent
\textbf{Verification by developers.}
Since there is no readily available ground truth for the warnings in the open-source repositories we studied, we reach out to core developers across all three repositories to obtain a ground truth on these warnings. We are also interested to see if developers agree with the labelling provided by \fs.

To achieve this, we first parse the git logs on the files containing the warnings we obtained from the three repositories. Specifically, we identify the developers who modified these warning lines the most along with their email IDs. 
We then send emails to three developers (one from each repository) and received responses from all three developers. The developers of tmux and openssh-portable agreed to collaborate with us. However the developer of openssl let us know that they would be unable to collaborate with us as it would take time away from their core development work. 

We provide each developer with the static analysis warning reports we obtained earlier from both RATS and Infer, and then ask them to review each static analysis warning and classify it as a True Positive (actual warning), False Positive (not a vulnerability), or ambiguous (if they are uncertain about it). 
We also give the developers the option to provide a reason for their decision.

It is worth noting that we opted not to send the warnings that were not compiled by our approach to the developers, as we recognized that such warnings would not contribute towards the analysis and could potentially waste the valuable time of open-source developers. Furthermore, we refrain from sharing the results obtained from \fs, as this could potentially bias the opinion of developers when classifying the warnings, and thereby introduce an unwanted element of subjectivity into the analysis.
Finally, it is important to note that for the openssh-portable project, the developer identified by our git log approach had invited another developer, as both were heavily involved in implementing the code flagged by these static analysis tools.
%

\begin{table}[tb!]
  \caption{Developer labels of warnings from the three open source repositories.}
  \centering
 \resizebox{\columnwidth}{!}{ 
 \begin{tabular}{l l r r r r}
    \toprule
    \multirow{2}{*}{Repository} & \multirow{2}{*}{\fs label} & \multirow{2}{*}{Total} & 
    \multicolumn{3}{c}{Developer label} \\ 
    & & & {FP}  & {TP} & {Ambiguous} \\
\midrule

\multirow{3}*{tmux} & PFP & 10 & 9 & 0 & 1 \\
& C& 3 & 1 & 0 & 2\\
& NR & 10 & 3 & 1 & 6\\

\hline
\multirow{3}*{openssh-portable}
& PFP & 24 & 24 & 0 & 0 \\
 & C & 3 & 3 & 0 & 0\\
& NR & 13 & 13 & 0 & 0\\

  \bottomrule
\end{tabular}
}
\label{devstats}
\end{table}
We present the results of our developer label analysis in Table~\ref{devstats}.
The table shows the labels assigned to the warnings by \fs (on the left side) and the labels assigned by the developers (on the top side).
As shown in Table~\ref{devstats}, the openssh-portable developers were able to confirm that all the possible false positives detected by \fs across both static analysis tools were indeed false positives. This outcome highlights the accuracy and reliability of our approach.
Table~\ref{devstats} also reveals that the developer from tmux was able to confirm that 9 out of the 10 possible false positives detected by \fs were indeed false positives.
However, for the remaining warnings, the developer labeled them as ambiguous because they reported that such cases require a precise call stack to analyze the warning further, as multiple callers can call the function.
Overall, our analysis shows that the developers' labels largely matched with the \fs classification of warnings as PFPs.


Table ~\ref{devstats} also reveals that the developers within the tmux repository labelled 9 warnings as ambiguous because they lacked sufficient context within large encryption-related functions.
Interestingly, of these 9 ambiguous warnings, \fs labelled 6 warnings as unreachable.
This is due to the inability of \fs to trigger the warnings within these large functions after 5 minutes of fuzzing, partially indicating agreement between ambiguous and not reachable warnings in the tmux repository.
Furthermore, the developer from tmux reported one warning with undefined behavior, which could potentially result in a buffer overflow. However, \fs was unable to execute that warning within tmux and classified it as not reachable.


 For most warnings that were labeled as false positives in tmux and openssh-portable, the developer supported their label with a rationale. 
 For example, the developers reported that in most false positives, the arguments passed to \textsc{libc} functions are within the correct bounds. Also, they mentioned that the index variable could not exceed buffer size within several loops. In other cases, bounds checking happens close to the warning, which prevents buffer overflow from occurring. 
 Finally, developers in openssh-portable also mentioned that they use third-party library calls from OpenBSD that are known to be completely safe.



\begin{table}[tb!]
  \caption{Evolution of possible false positives over time.}
  \centering
 \resizebox{0.95\columnwidth}{!}{ 
 \begin{tabular}{l r r}
    \toprule
    {Repository} & {\# PFP (3 years ago)} & {\# persistant PFP}\\
    \midrule
    openssl & 55 & 55 \\
    tmux & 11 & 8 \\
    openssh-portable & 20 & 20 \\
  \bottomrule
\end{tabular}
}
\label{groundtruth}
\end{table}


\conclusion{
    \textbf{Answer to RQ2:}
   We find that \fs is able to compile minimized slices for 244 out of 265 warnings in the three open-source repositories. In the tmux and openssh-portable repositories, \fs identified 33 out of 53 false positives confirmed by the project developers.
}

 \subsection{RQ3: Coverage, Warning Executions and Compilation of \fs}
 \label{res3}
  \begin{figure*}[tb!]
\centering
\includegraphics[width=.999\textwidth]{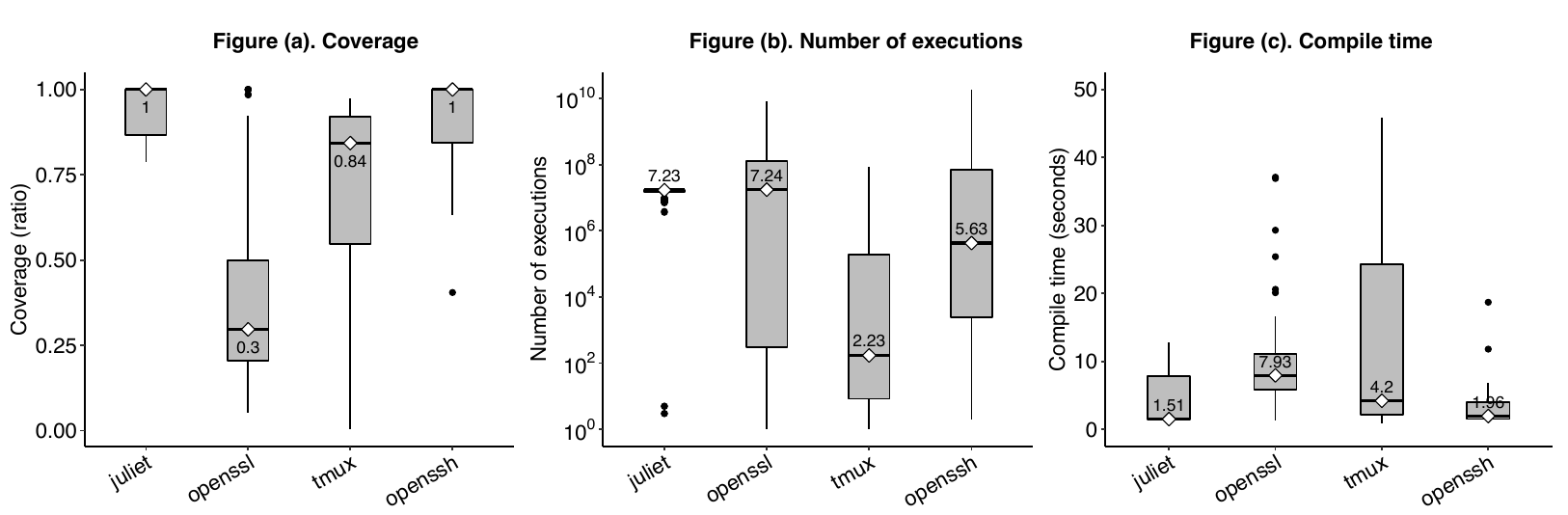}  
\caption{Coverage, number of executions on warning and compile time of minimized code slices.}
    \label{coverage}
\end{figure*}

Sections~\ref{res1} and~\ref{res2} presents
the overall effectiveness of \fs
with both synthetic and real-world case studies.
In this section,
we seek to provide a better understanding of \fs
from more fine-grained aspects.
Specifically, we examine \fs from three perspectives:
coverage,
number of executions on warning, and
compile time.\\
 


 \noindent
 \textbf{Coverage.}
 Code coverage is important to analyze because it indicates how much of the code within a given minimized code slice has been exercised during the fuzzing process.
 A higher code coverage means that more parts of the code have been exercised, which in turn increases the chances of confirming
 both true bugs and false alarms in static analysis reports.
 
We first evaluate the code coverage of the PFP code slices. 
Note that we obtain the coverage of the minimized slice in Step 5 of  Section~\ref{technique}. 
The ratio of the number of lines that are executed to the total number of lines is the code coverage percentage.
Figure~\ref{coverage}(a) shows the code slice coverage across the examined benchmarks. 
Note that this coverage is the coverage within the constructed code slice, not the coverage over the entire repository. 
Among the four datasets we examined, openssh and juliet showed the highest code coverage, with a median of 100\%. 
tmux was not far behind, with a median of 84\% code coverage. However, in the case of openssl, the median code coverage is close to 30\%, which is lower than other repositories. 
In fact, this is because some static analysis warnings in openssl are within functions that exercise whole modules within openssl, creating larger slices, which require extra time to achieve better coverage (recall that we limit the fuzzing time to five minutes).

\textcolor{black}{We also compute the slice size in the form of the total number of lines of code in the minimized slice (LOC). 
A higher slice size usually implies more dependencies needed to compile the function enclosing the warning.
Overall, we find that the size of slices varies depending on the examined project dataset and dependencies of the function enclosing the warning. 
For example, a minimized slice in the tmux project has a median of 7,691 LOC. However, the number is much lower for juliet, openssh, and openssl. The slice sizes (on median) in these projects are 58, 495, and 431, respectively.}\\

%

\noindent
\textbf{Frequency of warning hit.}
We analyze the frequency with which the warning line was executed for PFP.
By examining the number of executions for a warning line, we gain valuable insights into the likelihood that the warning may be a false positive. If a warning line is executed frequently without triggering a crash, this strongly suggests that it may not be indicative of an actual vulnerability. This is because the number of executions is closely tied to the diversity of input values evaluated by the fuzzer.
This information can be obtained through the coverage information
which is collected for fuzzing anyway.
This allows us to track the number of times
a given line of code is executed during testing.

 Figure~\ref{coverage}(b) presents the median number of executions for PFP warnings across each benchmark on a logarithmic scale. 
 Our analysis reveals that the median PFP warning was executed approximately 72.3 million times for juliet benchmark, 72.4 million times for openssl benchmark, 223 times for tmux benchmark, and 563 thousand times for openssh benchmark, respectively. 
 The executions on the warning vary depending on the warning location in the code, e.g., if they are surrounded by guard conditions, within for loops, etc.
 Our examination of code coverage and the number of executions on warning lines without crashes indicates that many of these warnings may indeed be false positives. The high levels of code coverage and the number of executions strongly indicate that these warnings do not correspond to actual vulnerabilities.\\
%
 %

\noindent
\textbf{Performance.}
To further assess the runtime performance of \fs, we conduct an evaluation of the time taken to generate the minimized code slices. Specifically, we measured the time required by \fs to create and compile each complete minimized slice for every warning.
Figure~\ref{coverage}(c) presents boxplots of the compile time in \fs for each benchmark. Our analysis reveals that code slices can be compiled within a range of 1.51-7.93 seconds. Notably, we did observe some outliers in the case of openssl, where functions belonging to multiple modules were searched and compiled, resulting in compile times of up to 80 seconds. However, overall, our results indicate that \fs is fast at constructing minimized code slices across all datasets.

Overall, these findings show that \fs is an effective tool for \textcolor{black}{pruning possible false positives} in static analysis warnings. By quickly generating minimized code slices, \fs can help developers and security professionals \textcolor{black}{prune} and mitigate PFP more efficiently.
\textcolor{black}{The process of constructing these slices takes on average less than 8 seconds, making \fs a valuable automatic approach for optimizing manual triage efforts.}

\conclusion{
   \textbf{Answer to RQ3:}
   \fs achieves a substantial coverage within the minimized slices of the  PFP warnings (median = 92.26\%).
   Also, \fs achieves a high number of executions on PFP warning lines (median = 69.87 million executions). In addition, \fs is able to compile minimized slices for the vast majority of the analyzed warnings in under 8 seconds.
}

\section{Retrospective analysis of PFP}
\label{Retrospective}
Although we were unable to obtain a ground truth for openssl from its developers, we can still evaluate the accuracy of \fs by analyzing the evolution of PFP detected over time. 
This technique was inspired by the work of Di Penta et al.~\cite{Penta2009Oct} and Aloraini et al.~\cite{Aloraini2019Dec}, who observed that warnings that persist in the same code for long periods without being removed are possible false positives.
The basic idea of the technique is that if a warning persists in the same code segment across multiple versions of the software and over a long time, then it is less likely to be a genuine vulnerability as it was not considered worth removing under any circumstances. 

To provide a comprehensive evaluation of our approach, we conduct an analysis not only on the openssl repository, but also on the two other open-source repositories included in our evaluation. 
Specifically, we consider a version of each repository that was 3 years old (i.e., the latest commit was made before January 1st, 2020), with the following git versions: openssl (5f95fbf399), tmux (566ab9aa), and openssh-portable (c4b2664b). 
We run RATS and Infer on these older versions of the repositories to flag buffer overflow warnings, and then use the \fs technique to \textcolor{black}{prune} all possible false positives (PFP) among these warnings. 
Next, we attempt to match these warnings with warnings in a more recent version of the repository. Specifically, we focus on the following two criteria: \textbf{(i)} the static analysis warning line in the older version (before 3 years) is identified as a possible false positive through \fs technique, and \textbf{(ii)} the warning line is still flagged by the respective static analysis tool in a recent version of the repository.
Table~\ref{groundtruth} presents the results of the evolution of the examined warnings.
The results indicate that for the openssl repository, all of the possible false positives are still flagged by the static analysis tool after 3 years. This suggests that there is a high likelihood that all of these warnings are indeed false positives.
Similarly, for the tmux repository, 8 out of 11 PFP are still flagged after 3 years.
In the case of openssh-portable, all 20 warnings matched both criteria. 
These findings support the efficacy of our approach in \textcolor{black}{pruning false positives} in static analysis warnings, and highlight the importance of considering the longevity and persistence of warnings in assessing their validity. 
\textcolor{black}{These results also indicate that \fs has the potential to help developers to deprioritize several such warnings for manual triage.}

 We found that in three cases, warnings that were identified as false positives in the 3-year-old version of the repository could not be matched to any warnings in the recent version. This occurred because, in two out of three cases, the code containing the warning had been deleted from the repository. The commits involved in deleting such code were not related to buffer overflow bugs. In the remaining case, the warning line had been modified, resulting in the static analysis warning being removed. This was due to a new feature in tmux that replaced internal representation of strings from UTF-8 to wide characters which modified the library call involved in this warning. 




\section{Limitations and Future Work}
\label{sec:futurework}

In Section~\ref{sec:results}, we report that we obtained 21 outlier cases that were not compiled. 
This was mainly due to the fact that \fs relies on srcML to parse C code. 
SrcML may misparse code and produce incorrect XML output. 
For example, srcML cannot correctly label code that contains inline assembly language within openssl. 
This is because srcML uses a grammar to parse the code and inline assembly language is not integrated into this grammar. 
As a result, errors in the srcML output can lead to required dependencies not being resolved and the slice not being created. Despite these limitations of srcML, it has been adopted by several previous works~\cite{Bui2021May,Bui2021Jul}.
 

%
\textcolor{black}{Also, we reported that 76 warnings were not reachable, meaning that the line could not be executed. This can be due to certain constraints on input, the requirement of external files, etc.}
Several of these cases, especially in openssl and openssh-portable, contain function pointers as arguments within the code slice. 
However, \fs currently does not fuzz function pointers. 
A possible way to address this limitation in future work is by first identifying all functions with matching signatures and return values that can be assigned to the function pointer.
Future work should also consider using a more efficient approach to refine indirect call targets~\cite{Lu2019Nov}.


Another limitation of \fs is related to fuzzing global variables.
Currently, \fs only provides a default initialization for global variables and does not mutate them, which led to the misclassification of 20 cases in the Juliet test suite as possible false positives.
However, despite not mutating function pointers and global variables, \fs is still capable of minimizing code that requires these components in their minimized slice.\\

\noindent
\textbf{Future work: fuzzing global variables.} 
In this work, we limit our techniques to provide a default initialization for global variables.
One potential extension for future work is to identify all global variables involved in the minimized slice and mutate them within the fuzzing wrapper in a similar way as function argument mutation.

\noindent
\textbf{Future work: finer-grained slicing.}
An interesting direction to explore would be to further
reduce the size of the program slice for fuzzing. In \fs, the entire function enclosing the vulnerability is considered for fuzzing. As future work, we plan to construct intra-function slices that minimally enclose the static analysis warning (both in terms of control-flow and data-flow). This can further ease the cost of fuzzing in pruning possible false positives.

\noindent
\textbf{Future work: supporting a diverse set of
static analysis tools.}
\textcolor{black}{Similar to related work~\cite{Bohme2017Oct,Woo2013Nov}, we illustrate \fs on a specific bug pattern: buffer overflow vulnerabilities. 
However, the concept of \fs can be extended to support static analysis tools that target different types of bugs, including but not limited to
integer overflow, null-pointer dereference,
use-after-free, dead code elimination, and even semantic and logic bugs.
To facilitate false alarm filtering on these types of bugs, we only need to replace the oracle that detects the violation at runtime (e.g., UBSAN is an oracle for integer overflows~\cite{UBSAN}) within the \fs framework.}




\section{Related work}

A common approach to fuzzing has been to fuzz independent submodules, drivers or libraries separately. 
There is a rich literature focused on fuzzing independent libraries such as Transport Layer Security (TLS), deep learning, C/C++ libraries~\cite{Wei2022May,Jang2019Dec,Somorovsky2016Oct}.
For example, Corina et al.~\cite{Corina2017Oct} proposed fuzzing for kernel drivers effectively finding bugs within them. 
\fs differs from such approaches since it involves fuzzing at the function level for any arbitrary warning location, aiming to \textcolor{black}{prune possible false positives.}

There exists a rich literature on directing fuzzing towards a given location~\cite{272157,Huang2022May,Wustholz2019May,Chen2018Oct,Bohme2017Oct, Razzer}. The core idea behind such methods is to mutate inputs that are closer to reaching the target location.
The main difference between \fs and directed fuzzing is that
\fs does not use \verb|main| method as the entrypoint. Instead, \fs creates the minimal slice enclosing the warning first and then we confine the state space exploration within the slice. In fact, their techniques to direct input mutation towards a certain location are orthogonal to our approach and can be used as a complementary technique within \fs.

Fuzzing has been applied to binary-level code slices as well. For instance, Chen et al.~\cite{Chen2022Nov} implemented fuzzing on independent code snippets extracted from real-time operating systems (RTOS) binaries. 
In contrast, \fs takes a different approach by creating code slices at the source code level instead of the binary level. By generating fully compiled slices, \fs identifies possible false positives in static analysis reports, eliminating the need to deal with unstable fuzzing caused by incomplete context.


The work closest to \fs utilizes symbolic execution.
For example, Engler et al.~\cite{engler2007under}
proposed the idea of under-constrained symbolic execution (UC-KLEE).
UC-KLEE takes an arbitrary function and
symbolically executes it without initializing
any of its data structures and
without doing any environment setup,
with the goal of finding quality bugs
in drivers within the Linux operating system.
While program slicing helps to reduce the search space of
symbolic execution, UC-KLEE still suffers from other sources
of path explosion and imprecision such as
unbounded loops,
pointer arithmetic,
memory modeling, and
invocation of library functions.
\fs, on the other hand,
is a dynamic analysis tool at its core and
does not suffer from the above-mentioned limitations.
Instead, \fs is subject to a different set of limitations such as code coverage and
effectiveness of mutation strategies.

\textcolor{black}{Kallingal et al.~\cite{TPSlice} generate code slices aimed at identifying true bugs through the Helium framework. Helium can work only with certain static analysis tools - those that can provide a list of several statements leading to a given warning. In the Helium approach, the slice considered is a least common ancestor subtree of the parse tree over the full path leading to the warning. \fs on the other hand does not make any assumptions about the static analysis tool and does not rely on the accuracy of the static analysis tool in reporting paths leading to a warning. Additionally, since \fs aims to only prune false positives, the framework can disregard most of the program and directly fuzz only the function enclosing the warning. Furthermore, Kallingal et al. mention they are able to compile 68.5\% of their code fragments. Since \fs compiles a smaller slice containing the enclosing function and its dependencies it compiled 244 out of 265 slices (92\%) in open-source repositories and all the slices (100\%) in the Juliet test suite. Such statistics from both works highlight the difficulty of generating compilable slices relevant to a warning and prove that the task is not trivial. 
}

\textcolor{black}{A plethora of work proposed approaches that utilized machine learning to reduce false positives in static analysis tools~\cite{Yoon2014Dec,Koc2017Jun,ML1,ML2}.
Hanam et al.~\cite{ML1} create a feature vector based on code characteristics at the site of each warning. The technique leverages machine learning techniques to build an actionable alert prediction model. 
Yedida et al.~\cite{ML2} proposed locally adjusting decision boundaries of models for actionable warnings to improve overall performance. 
\fs differs from such machine learning-based works since it actually dynamically executes the program to classify the warning.}

\textcolor{black}{Recent techniques such as~\cite{Driver1,Driver2,Driver3,Driver4} focused on improvements in automatic fuzzing driver generation, which is orthogonal to the \fs approach. \fs can benefit from the recent state of the art in this area. Tip et al.~\cite{Tip1994Jul} surveyed algorithmic aspects of program slicing techniques. However, in \fs, the slice must also be compiled and linked into an executable with the correct compiler options, which requires storing relevant information of the build system. This increases the complexity of the slicing component within the \fs framework.
}
\section{Conclusion}
This paper introduces \fs, a framework that automates the pruning of false positives from static analysis tool warnings. We achieve this by fuzzing warnings at the function level, as it identifies non-crashing fuzzed warnings as potential false positives. The framework employs two steps: (1) creating a minimal compiled code slice containing any warning and (2) generating a fuzzing wrapper that performs type-based input generation for the enclosing function. 
Evaluation on synthetic and real-world C codebases demonstrates FuzzSlice's effectiveness.
In the synthetic Juliet test suite, \fs identifies all 864 false positives (100\%). In open-source repositories (tmux and openssh-portable), where developers independently labeled warnings from the static analysis tools, \fs identifies 33 potential false positives out of the 53 confirmed by developers (62.2\%). 
Thus, \fs substantially reduces the effort required for developers to examine warnings. 
Additionally, in the Juliet test suite, 20 of the 884 possible false positives detected by \fs were actually true positives (2.2\%). We were able to confirm that the incorrect classification in the Juliet test suite was due to our inability to fuzz global variables.
In tmux and openssh-portable, of the 34 warnings we determined as possible false positives, 33 were confirmed as false positives by the developers, and 1 case was deemed as ambiguous (2.9\%). These results validate the key insight of the \fs framework that a warning that does not yield a crash when fuzzed at the function level in a given time budget is a possible false positive.

\section*{Acknowledgment}
We would like to thank the developers of the analyzed open source repositories in this paper - Damien Miller (openssh-portable), Darren Tucker (openssh-portable) and Nicholas Marriott (tmux) for reviewing the warnings used in the analysis. Also, we thank NSERC who partially supported the work in this paper through funding from Alliance and WHJIL.

\balance
\bibliographystyle{IEEEtran}
\bibliography{Bibliography}

\begin{thebibliography}{10}
\providecommand{\url}[1]{#1}
\csname url@samestyle\endcsname
\providecommand{\newblock}{\relax}
\providecommand{\bibinfo}[2]{#2}
\providecommand{\BIBentrySTDinterwordspacing}{\spaceskip=0pt\relax}
\providecommand{\BIBentryALTinterwordstretchfactor}{4}
\providecommand{\BIBentryALTinterwordspacing}{\spaceskip=\fontdimen2\font plus
\BIBentryALTinterwordstretchfactor\fontdimen3\font minus \fontdimen4\font\relax}
\providecommand{\BIBforeignlanguage}[2]{{%
\expandafter\ifx\csname l@#1\endcsname\relax
\typeout{** WARNING: IEEEtran.bst: No hyphenation pattern has been}%
\typeout{** loaded for the language `#1'. Using the pattern for}%
\typeout{** the default language instead.}%
\else
\language=\csname l@#1\endcsname
\fi
#2}}
\providecommand{\BIBdecl}{\relax}
\BIBdecl

\bibitem{Johnson}
B.~Johnson, Y.~Song, E.~Murphy-Hill, and R.~Bowdidge, ``{Why don't software developers use static analysis tools to find bugs?}'' in \emph{{2013 35th International Conference on Software Engineering (ICSE)}}.\hskip 1em plus 0.5em minus 0.4em\relax IEEE, pp. 18--26.

\bibitem{alfadel2023discoverability}
M.~Alfadel, D.~E. Costa, E.~Shihab, and B.~Adams, ``On the discoverability of npm vulnerabilities in node. js projects,'' \emph{ACM Transactions on Software Engineering and Methodology}, vol.~32, no.~4, pp. 1--27, 2023.

\bibitem{Cheirdari}
F.~Cheirdari and G.~Karabatis, ``{Analyzing False Positive Source Code Vulnerabilities Using Static Analysis Tools},'' in \emph{{2018 IEEE International Conference on Big Data (Big Data)}}.\hskip 1em plus 0.5em minus 0.4em\relax IEEE, pp. 10--13.

\bibitem{Nadeem2012Mar}
M.~Nadeem, B.~J. Williams, and E.~B. Allen, ``{High false positive detection of security vulnerabilities: a case study},'' in \emph{{ACM-SE '12: Proceedings of the 50th Annual Southeast Regional Conference}}.\hskip 1em plus 0.5em minus 0.4em\relax New York, NY, USA: Association for Computing Machinery, Mar. 2012, pp. 359--360.

\bibitem{Kang2022May}
H.~J. Kang, K.~L. Aw, and D.~Lo, ``{Detecting false alarms from automatic static analysis tools: how far are we?}'' in \emph{{ICSE '22: Proceedings of the 44th International Conference on Software Engineering}}.\hskip 1em plus 0.5em minus 0.4em\relax New York, NY, USA: Association for Computing Machinery, May 2022, pp. 698--709.

\bibitem{Park2016May}
\BIBentryALTinterwordspacing
J.~Park, I.~Lim, and S.~Ryu, ``{Battles with False Positives in Static Analysis of JavaScript Web Applications in the Wild},'' in \emph{{2016 IEEE/ACM 38th International Conference on Software Engineering Companion (ICSE-C)}}.\hskip 1em plus 0.5em minus 0.4em\relax IEEE, May 2016, pp. 61--70. [Online]. Available: \url{https://ieeexplore.ieee.org/document/7883289}
\BIBentrySTDinterwordspacing

\bibitem{Aloraini2017Sep}
B.~Aloraini and M.~Nagappan, ``{Evaluating State-of-the-Art Free and Open Source Static Analysis Tools Against Buffer Errors in Android Apps},'' in \emph{{2017 IEEE International Conference on Software Maintenance and Evolution (ICSME)}}.\hskip 1em plus 0.5em minus 0.4em\relax IEEE, Sep. 2017, pp. 295--306.

\bibitem{Christakis2016Aug}
M.~Christakis and C.~Bird, ``{What developers want and need from program analysis: an empirical study},'' in \emph{{ASE '16: Proceedings of the 31st IEEE/ACM International Conference on Automated Software Engineering}}.\hskip 1em plus 0.5em minus 0.4em\relax New York, NY, USA: Association for Computing Machinery, Aug. 2016, pp. 332--343.

\bibitem{Johnson2013May}
B.~Johnson, Y.~Song, E.~Murphy-Hill, and R.~Bowdidge, ``{Why don't software developers use static analysis tools to find bugs?}'' in \emph{{2013 35th International Conference on Software Engineering (ICSE)}}.\hskip 1em plus 0.5em minus 0.4em\relax IEEE, May 2013, pp. 672--681.

\bibitem{Bohme2017Oct}
M.~B{\ifmmode\ddot{o}\else\"{o}\fi}hme, V.-T. Pham, M.-D. Nguyen, and A.~Roychoudhury, ``{Directed Greybox Fuzzing},'' in \emph{{CCS '17: Proceedings of the 2017 ACM SIGSAC Conference on Computer and Communications Security}}.\hskip 1em plus 0.5em minus 0.4em\relax New York, NY, USA: Association for Computing Machinery, Oct. 2017, pp. 2329--2344.

\bibitem{Christakis2016May}
M.~Christakis, P.~M{\ifmmode\ddot{u}\else\"{u}\fi}ller, and V.~W{\ifmmode\ddot{u}\else\"{u}\fi}stholz, ``{Guiding dynamic symbolic execution toward unverified program executions},'' in \emph{{ICSE '16: Proceedings of the 38th International Conference on Software Engineering}}.\hskip 1em plus 0.5em minus 0.4em\relax New York, NY, USA: Association for Computing Machinery, May 2016, pp. 144--155.

\bibitem{Bohme2020Nov}
M.~B{\ifmmode\ddot{o}\else\"{o}\fi}hme and B.~Falk, ``{Fuzzing: on the exponential cost of vulnerability discovery},'' in \emph{{ESEC/FSE 2020: Proceedings of the 28th ACM Joint Meeting on European Software Engineering Conference and Symposium on the Foundations of Software Engineering}}.\hskip 1em plus 0.5em minus 0.4em\relax New York, NY, USA: Association for Computing Machinery, Nov. 2020, pp. 713--724.

\bibitem{Taintscope}
T.~Wang, T.~Wei, G.~Gu, and W.~Zou, ``{TaintScope: A Checksum-Aware Directed Fuzzing Tool for Automatic Software Vulnerability Detection},'' in \emph{{2010 IEEE Symposium on Security and Privacy}}.\hskip 1em plus 0.5em minus 0.4em\relax IEEE, pp. 16--19.

\bibitem{272157}
\BIBentryALTinterwordspacing
G.~Lee, W.~Shim, and B.~Lee, ``Constraint-guided directed greybox fuzzing,'' in \emph{30th USENIX Security Symposium (USENIX Security 21)}.\hskip 1em plus 0.5em minus 0.4em\relax USENIX Association, Aug. 2021, pp. 3559--3576. [Online]. Available: \url{https://www.usenix.org/conference/usenixsecurity21/presentation/lee-gwangmu}
\BIBentrySTDinterwordspacing

\bibitem{defuzz}
X.~Zhu, S.~Liu, X.~Li, S.~Wen, J.~Zhang, C.~Seyit, and Y.~Xiang, ``{DeFuzz: Deep Learning Guided Directed Fuzzing},'' \emph{arXiv}, Oct. 2020.

\bibitem{Wei2022May}
A.~Wei, Y.~Deng, C.~Yang, and L.~Zhang, ``{Free lunch for testing: fuzzing deep-learning libraries from open source},'' in \emph{{ICSE '22: Proceedings of the 44th International Conference on Software Engineering}}.\hskip 1em plus 0.5em minus 0.4em\relax New York, NY, USA: Association for Computing Machinery, May 2022, pp. 995--1007.

\bibitem{Jang2019Dec}
J.~Jang and H.~K. Kim, ``{FuzzBuilder: automated building greybox fuzzing environment for C/C++ library},'' in \emph{{ACSAC '19: Proceedings of the 35th Annual Computer Security Applications Conference}}.\hskip 1em plus 0.5em minus 0.4em\relax New York, NY, USA: Association for Computing Machinery, Dec. 2019, pp. 627--637.

\bibitem{Somorovsky2016Oct}
J.~Somorovsky, ``{Systematic Fuzzing and Testing of TLS Libraries},'' in \emph{{CCS '16: Proceedings of the 2016 ACM SIGSAC Conference on Computer and Communications Security}}.\hskip 1em plus 0.5em minus 0.4em\relax New York, NY, USA: Association for Computing Machinery, Oct. 2016, pp. 1492--1504.

\bibitem{Chen2022Nov}
L.~Chen, Q.~Cai, Z.~Ma, Y.~Wang, H.~Hu, M.~Shen, Y.~Liu, S.~Guo, H.~Duan, K.~Jiang, and Z.~Xue, ``{SFuzz: Slice-based Fuzzing for Real-Time Operating Systems},'' in \emph{{CCS '22: Proceedings of the 2022 ACM SIGSAC Conference on Computer and Communications Security}}.\hskip 1em plus 0.5em minus 0.4em\relax New York, NY, USA: Association for Computing Machinery, Nov. 2022, pp. 485--498.

\bibitem{openssl2023Jan}
\BIBentryALTinterwordspacing
openssl, ``{openssl},'' Jan. 2023, [Online; accessed 29. Jan. 2023]. [Online]. Available: \url{https://github.com/openssl/openssl}
\BIBentrySTDinterwordspacing

\bibitem{andrew-d.2023Jan}
\BIBentryALTinterwordspacing
andrew d., ``{rough-auditing-tool-for-security},'' Jan. 2023, [Online; accessed 29. Jan. 2023]. [Online]. Available: \url{https://github.com/andrew-d/rough-auditing-tool-for-security}
\BIBentrySTDinterwordspacing

\bibitem{exampleloc}
\BIBentryALTinterwordspacing
openssl, ``{openssl},'' Jan. 2023, [Online; accessed 29. Jan. 2023]. [Online]. Available: \url{https://github.com/openssl/openssl/blob/master/test/testutil/driver.c}
\BIBentrySTDinterwordspacing

\bibitem{BibEntry2023Mar}
\BIBentryALTinterwordspacing
``{american fuzzy lop},'' Mar. 2023, [Online; accessed 10. Mar. 2023]. [Online]. Available: \url{https://lcamtuf.coredump.cx/afl}
\BIBentrySTDinterwordspacing

\bibitem{BibEntry2023Jan}
\BIBentryALTinterwordspacing
``{libFuzzer {\textendash} a library for coverage-guided fuzz testing. {\ifmmode---\else\textemdash\fi} LLVM 17.0.0git documentation},'' Jan. 2023, [Online; accessed 29. Jan. 2023]. [Online]. Available: \url{https://llvm.org/docs/LibFuzzer.html}
\BIBentrySTDinterwordspacing

\bibitem{Calcagno2011}
C.~Calcagno and D.~Distefano, ``{Infer: An Automatic Program Verifier for Memory Safety of C Programs},'' in \emph{{NASA Formal Methods}}.\hskip 1em plus 0.5em minus 0.4em\relax Berlin, Germany: Springer, 2011, pp. 459--465.

\bibitem{rizsotto2023Jan}
\BIBentryALTinterwordspacing
rizsotto, ``{Bear},'' Jan. 2023, [Online; accessed 30. Jan. 2023]. [Online]. Available: \url{https://github.com/rizsotto/Bear}
\BIBentrySTDinterwordspacing

\bibitem{Collard}
M.~L. Collard, M.~J. Decker, and J.~I. Maletic, ``{srcML: An Infrastructure for the Exploration, Analysis, and Manipulation of Source Code: A Tool Demonstration},'' in \emph{{2013 IEEE International Conference on Software Maintenance}}.\hskip 1em plus 0.5em minus 0.4em\relax IEEE, pp. 22--28.

\bibitem{stallman1988debugging}
R.~Stallman, R.~Pesch, S.~Shebs \emph{et~al.}, ``Debugging with gdb,'' \emph{Free Software Foundation}, vol. 675, 1988.

\bibitem{serebryany2012addresssanitizer}
K.~Serebryany, D.~Bruening, A.~Potapenko, and D.~Vyukov, ``Addresssanitizer: A fast address sanity checker,'' 2012.

\bibitem{Boland2012Oct}
T.~Boland and P.~E. Black, ``{Juliet 1.1 C/C++ and Java Test Suite},'' \emph{Computer}, vol.~45, no.~10, pp. 88--90, Oct. 2012.

\bibitem{openssh2023Jan}
\BIBentryALTinterwordspacing
openssh, ``{openssh-portable},'' Jan. 2023, [Online; accessed 30. Jan. 2023]. [Online]. Available: \url{https://github.com/openssh/openssh-portable}
\BIBentrySTDinterwordspacing

\bibitem{tmux2023Jan}
\BIBentryALTinterwordspacing
tmux, ``{tmux},'' Jan. 2023, [Online; accessed 30. Jan. 2023]. [Online]. Available: \url{https://github.com/tmux/tmux}
\BIBentrySTDinterwordspacing

\bibitem{Penta2009Oct}
M.~D. Penta, L.~Cerulo, and L.~Aversano, ``{The life and death of statically detected vulnerabilities: An empirical study},'' \emph{Information and Software Technology}, vol.~51, no.~10, pp. 1469--1484, Oct. 2009.

\bibitem{Aloraini2019Dec}
B.~Aloraini, M.~Nagappan, D.~M. German, S.~Hayashi, and Y.~Higo, ``{An empirical study of security warnings from static application security testing tools},'' \emph{Journal of Systems and Software}, vol. 158, p. 110427, Dec. 2019.

\bibitem{Bui2021May}
N.~D.~Q. Bui, Y.~Yu, and L.~Jiang, ``{InferCode: Self-Supervised Learning of Code Representations by Predicting Subtrees},'' in \emph{{2021 IEEE/ACM 43rd International Conference on Software Engineering (ICSE)}}.\hskip 1em plus 0.5em minus 0.4em\relax IEEE, May 2021, pp. 1186--1197.

\bibitem{Bui2021Jul}
------, ``{Self-Supervised Contrastive Learning for Code Retrieval and Summarization via Semantic-Preserving Transformations},'' in \emph{{SIGIR '21: Proceedings of the 44th International ACM SIGIR Conference on Research and Development in Information Retrieval}}.\hskip 1em plus 0.5em minus 0.4em\relax New York, NY, USA: Association for Computing Machinery, Jul. 2021, pp. 511--521.

\bibitem{Lu2019Nov}
K.~Lu and H.~Hu, ``{Where Does It Go? Refining Indirect-Call Targets with Multi-Layer Type Analysis},'' in \emph{{CCS '19: Proceedings of the 2019 ACM SIGSAC Conference on Computer and Communications Security}}.\hskip 1em plus 0.5em minus 0.4em\relax New York, NY, USA: Association for Computing Machinery, Nov. 2019, pp. 1867--1881.

\bibitem{Woo2013Nov}
M.~Woo, S.~K. Cha, S.~Gottlieb, and D.~Brumley, ``{Scheduling black-box mutational fuzzing},'' in \emph{{CCS '13: Proceedings of the 2013 ACM SIGSAC conference on Computer {\&} communications security}}.\hskip 1em plus 0.5em minus 0.4em\relax New York, NY, USA: Association for Computing Machinery, Nov. 2013, pp. 511--522.

\bibitem{UBSAN}
\BIBentryALTinterwordspacing
``{UndefinedBehaviorSanitizer {\ifmmode---\else\textemdash\fi} Clang 17.0.0git documentation},'' Jun. 2023, [Online; accessed 27. Jun. 2023]. [Online]. Available: \url{https://clang.llvm.org/docs/UndefinedBehaviorSanitizer.html}
\BIBentrySTDinterwordspacing

\bibitem{Corina2017Oct}
J.~Corina, A.~Machiry, C.~Salls, Y.~Shoshitaishvili, S.~Hao, C.~Kruegel, and G.~Vigna, ``{DIFUZE: Interface Aware Fuzzing for Kernel Drivers},'' in \emph{{CCS '17: Proceedings of the 2017 ACM SIGSAC Conference on Computer and Communications Security}}.\hskip 1em plus 0.5em minus 0.4em\relax New York, NY, USA: Association for Computing Machinery, Oct. 2017, pp. 2123--2138.

\bibitem{Huang2022May}
H.~Huang, Y.~Guo, Q.~Shi, P.~Yao, R.~Wu, and C.~Zhang, ``{BEACON: Directed Grey-Box Fuzzing with Provable Path Pruning},'' in \emph{{2022 IEEE Symposium on Security and Privacy (SP)}}.\hskip 1em plus 0.5em minus 0.4em\relax IEEE, May 2022, pp. 36--50.

\bibitem{Wustholz2019May}
V.~W{\ifmmode\ddot{u}\else\"{u}\fi}stholz and M.~Christakis, ``{Targeted Greybox Fuzzing with Static Lookahead Analysis},'' \emph{arXiv}, May 2019.

\bibitem{Chen2018Oct}
H.~Chen, Y.~Xue, Y.~Li, B.~Chen, X.~Xie, X.~Wu, and Y.~Liu, ``{Hawkeye: Towards a Desired Directed Grey-box Fuzzer},'' in \emph{{CCS '18: Proceedings of the 2018 ACM SIGSAC Conference on Computer and Communications Security}}.\hskip 1em plus 0.5em minus 0.4em\relax New York, NY, USA: Association for Computing Machinery, Oct. 2018, pp. 2095--2108.

\bibitem{Razzer}
D.~R. Jeong, K.~Kim, B.~Shivakumar, B.~Lee, and I.~Shin, ``{Razzer: Finding Kernel Race Bugs through Fuzzing},'' in \emph{{2019 IEEE Symposium on Security and Privacy (SP)}}.\hskip 1em plus 0.5em minus 0.4em\relax IEEE, May 2019, pp. 754--768.

\bibitem{engler2007under}
D.~Engler and D.~Dunbar, ``Under-constrained execution: making automatic code destruction easy and scalable,'' in \emph{Proceedings of the 2007 international symposium on Software Testing and analysis}, 2007, pp. 1--4.

\bibitem{TPSlice}
A.~Kallingal~Joshy, X.~Chen, B.~Steenhoek, and W.~Le, ``{Validating static warnings via testing code fragments},'' in \emph{{ISSTA 2021: Proceedings of the 30th ACM SIGSOFT International Symposium on Software Testing and Analysis}}.\hskip 1em plus 0.5em minus 0.4em\relax New York, NY, USA: Association for Computing Machinery, Jul. 2021, pp. 540--552.

\bibitem{Yoon2014Dec}
J.~Yoon, M.~Jin, and Y.~Jung, ``{Reducing False Alarms from an Industrial-Strength Static Analyzer by SVM},'' in \emph{{APSEC '14: Proceedings of the 2014 21st Asia-Pacific Software Engineering Conference - Volume 02}}.\hskip 1em plus 0.5em minus 0.4em\relax USA: IEEE Computer Society, Dec. 2014, pp. 3--6.

\bibitem{Koc2017Jun}
U.~Koc, P.~Saadatpanah, J.~S. Foster, and A.~A. Porter, ``{Learning a classifier for false positive error reports emitted by static code analysis tools},'' in \emph{{MAPL 2017: Proceedings of the 1st ACM SIGPLAN International Workshop on Machine Learning and Programming Languages}}.\hskip 1em plus 0.5em minus 0.4em\relax New York, NY, USA: Association for Computing Machinery, Jun. 2017, pp. 35--42.

\bibitem{ML1}
Q.~Hanam, L.~Tan, R.~Holmes, and P.~Lam, ``{Finding patterns in static analysis alerts: improving actionable alert ranking},'' in \emph{{MSR 2014: Proceedings of the 11th Working Conference on Mining Software Repositories}}.\hskip 1em plus 0.5em minus 0.4em\relax New York, NY, USA: Association for Computing Machinery, May 2014, pp. 152--161.

\bibitem{ML2}
R.~Yedida, H.~J. Kang, H.~Tu, X.~Yang, D.~Lo, and T.~Menzies, ``{How to Find Actionable Static Analysis Warnings: A Case Study With FindBugs},'' \emph{IEEE Trans. Software Eng.}, vol.~49, no.~4, pp. 2856--2872, Jan. 2023.

\bibitem{Driver1}
\BIBentryALTinterwordspacing
D.~Babic, S.~Bucur, Y.~Chen, F.~Ivancic, T.~King, M.~Kusano, C.~Lemieux, L.~Szekeres, and W.~Wang, ``{FUDGE: Fuzz Driver Generation at Scale},'' \emph{Google Research}, 2019. [Online]. Available: \url{https://research.google/pubs/pub48314}
\BIBentrySTDinterwordspacing

\bibitem{Driver2}
M.~Zhang, J.~Liu, F.~Ma, H.~Zhang, and Y.~Jiang, ``{IntelliGen: automatic driver synthesis for fuzz testing},'' in \emph{{ICSE-SEIP '21: Proceedings of the 43rd International Conference on Software Engineering: Software Engineering in Practice}}.\hskip 1em plus 0.5em minus 0.4em\relax IEEE Press, May 2021, pp. 318--327.

\bibitem{Driver3}
K.~K. Ispoglou, D.~Austin, V.~Mohan, and M.~Payer, ``{FuzzGen: automatic fuzzer generation},'' in \emph{{SEC'20: Proceedings of the 29th USENIX Conference on Security Symposium}}.\hskip 1em plus 0.5em minus 0.4em\relax USA: USENIX Association, Aug. 2020, pp. 2271--2287.

\bibitem{Driver4}
G.~Fraser and A.~Arcuri, ``{EvoSuite: automatic test suite generation for object-oriented software},'' in \emph{{ESEC/FSE '11: Proceedings of the 19th ACM SIGSOFT symposium and the 13th European conference on Foundations of software engineering}}.\hskip 1em plus 0.5em minus 0.4em\relax New York, NY, USA: Association for Computing Machinery, Sep. 2011, pp. 416--419.

\bibitem{Tip1994Jul}
F.~Tip, \emph{{A Survey of Program Slicing Techniques.}}\hskip 1em plus 0.5em minus 0.4em\relax CWI (Centre for Mathematics and Computer Science), Jul. 1994.

\end{thebibliography}

\end{document}